\documentclass[article,nofootinbib]{revtex4-1}
\usepackage{fullpage,amsmath,physics,hyperref}
\usepackage{graphicx}

\begin{document}
\title{Fluctuations in the stress energy tensor of spinor fields evolving in general FRW spacetimes}
\author{Ankit Dhanuka}
\email{ankitdhanuka555@gmail.com, ph17006@iisermohali.ac.in}
\affiliation{Department of Physical Sciences, IISER Mohali, \\ Sector 81, SAS Nagar, Manauli PO 140306, Punjab, India}
\begin{abstract}
In this work, we study quantum fluctuations in the stress energy tensor of spinor fields evolving in general FRW spacetimes. We quantify these fluctuations by the noise kernel of spinor fields. For the particular case of de Sitter spacetime, we place the spinor field in what is called the fermionic Bunch Davies vacuum and study the variation of the noise kernel with the mass of the field. We then make use of the conformal invariance of massless spinor fields and employ an equivalence that relates a massless spinor field in any given FRW spacetime with a corresponding massless spinor field in de Sitter spacetime. Using this equivalence, we study the behaviour of the noise kernel of massless spinor fields in general FRW spacetimes while placing the fields in the Bunch Davies vacuum of the corresponding massless spinor field of the de Sitter spacetime. We extend this analysis to the coordinate invariant quantity of the energy energy correlator. We compare these results for spinor fields in considered spacetimes with the analogous results for scalar fields. This study helps us better understand whether the spin $1/2$ fermionic matter remains strongly correlated in considered spacetimes leading to considerable backreaction or not. 
\end{abstract}
\maketitle
\section{Introduction}
With the ever so precise cosmological data \cite{Planck:2018vyg,Planck:2019kim,Planck:2013win} becoming available to us, we have increasing density of empirical references against which we can test the validity of our theories, arrived at either through intuition or to explain observations obtained in other astrophysical or cosmological contexts, and gain new insights into the origin of the universe and its evolution history. For example, the anisotropies in the cosmic microwave background (CMB) are believed to be related to the power spectrum of the quantum fluctuations of the inflaton field \cite{Weinberg:2008zzc,Baumann:2009ds} and therefore, corrections to this quantity coming from higher order effects can show up themselves in the finer details of the observed anisotropies \cite{Bartolo:2004if}. In scenarios where fluctuations over the background quantities can be considered small, we see that the matter fluctuations (over the background matter fields) can couple to the gravitational fluctuations (over the background metric) through their stress energy tensors\footnote{This can be easily seen by noticing that 
\begin{eqnarray}
S_{total} &=& S_{gravity}(\tilde{g}) + S_{matter}(\tilde{g},\tilde{\phi} ) \nonumber \\ &=& S_{gravity}(g_{bg}+h) + S_{matter}(g_{bg}+h,\phi_{bg} + \delta\phi) \nonumber\\ &=& \Big(S^{(0)}_{gravity}(g_{bg}) + S^{(0)}_{matter}(g_{bg},\phi_{bg})\Big) + S^{(2)}_{matter}(g_{bg},\delta\phi) + S^{(2)}_{gravity}(h) \nonumber \\ & &  \hspace{2cm} + \int d^4x \sqrt{-g_{bg}}\Big(-\frac{1}{2}h_{ab}T^{ab}_{\delta\phi}\Big) + .... \nonumber
\end{eqnarray} 
where $g_{bg}$ and $\phi_{bg}$ are the background metric and the background matter field respectively and similarly $h$ and $\delta\phi$ are the respective fluctuations over these background quantities. We have dropped the terms which are first order in $h$ or $\delta\phi$ as they contain expressions which correspond to the equations of motion for the background quantities and, hence, are zero. Here the terms $S^{(2)}$ are second order in $h$ and $\delta \phi$ and provide the free dynamics of the matter and the gravity fluctuations. We see that one of the interaction terms between the gravity fluctuations and the matter field fluctuations is proportional to the stress energy tensor of the matter field fluctuations i.e., $T^{ab}_{\delta\phi} = -\frac{2}{\sqrt{-g_{bg}}}\frac{\delta S_{matter}(\delta\phi)}{\delta g_{ab}(bg)}$.}. Therefore, if we quantize these fluctuations, then such an interaction term can be seen as providing a perturbed Hamiltonian term to the free dynamics of these fluctuations and as such the propagators of these fluctuations get corrections from this term \cite{Weinberg:2005vy}. From the structure of the interaction term, it is clear that it would inevitably involve n-point correlators of the matter field stress energy operators. Calculating n-point correlators of stress-energy operators is, in general, a very difficult task. However, in order to study the 2nd order effect (after taking into account the first order semiclassical effects of the 1-point quantum averages of stress energy operators) coming from this interaction term, one only needs the knowledge of two point correlators of quantum field stress energy operators. A number of previous investigations have undertaken the task of calculating and studying the behaviour of these two point correlations of quantum field stress energy operators (mainly, that of scalar fields). These works have considered the dynamics of quantum fields over different types of background spacetimes ranging from the Minkowski spacetime \cite{Ford:2005sp,Frob:2013sxa,Bates:2013lya} to the de Sitter spacetime \cite{Perez-Nadal:2009jcz,Dhanuka:2020yxp} to the more general conformally flat spacetimes \cite{Bates:2013iq} among others \cite{Eftekharzadeh:2010qp,Satin:2012wj,Wu:2021ktn}. These studies have been performed by placing quantum fields in different types of vacua, for example, \cite{Perez-Nadal:2009jcz,Dhanuka:2020yxp} have carried out the analysis by placing the scalar field in the Bunch Davies vacuum \cite{Bunch:1978yq,Allen:1985ux}. A number of works have also focussed attention on how the fluctuations in quantum stress energy tensors affect the geometry of spacetimes, for example \cite{Borgman:2003dm,Ford:2010wd} have considered how congruences of geodesics are affected by these fluctuations and similarly, \cite{Wu:2006ew,Wu:2011gk,Hsiang:2017top} have considered how these fluctuations manifest themselves in the power spectra of primordial gravity waves and hence provide possible observable effects of these fluctuations. These fluctuations in stress energy operators, or more appropriately the noise kernel (which is the vacuum expectation of symmetrized stress energy bi-tensors, see below), play a fundamental role in the stochastic gravity \cite{Hu:2008rga} framework where they feed the stochastic fluctuations appearing in the Einstein Langevin equations. Therefore, in any attempt to go beyond the first order semiclassical analyses, we would always end up requiring to calculate the behaviour of the noise kernel of matter fields. \\
Since the present-day universe contains both bosonic and fermionic types of particles, it is natural to assume that (during their journey through the evolution of the universe) they could couple with gravitational waves with the above types of interaction terms. Though majority of the above cited works have considered only the effects of bosonic fields such as scalar and electromagnetic fields, there are some works that have also considered the effects of fermionic fields on gravitational waves. For example, assuming that the fermions were produced after inflation, in the reheating era \cite{Peloso:2000hy,Giudice:1999fb}, \cite{Figueroa:2013vif} has considered the backreaction of fermionic fields on gravitational waves. Similarly, \cite{Feng:2012jm,Feng:2015dxa} have looked at the modification in gravitational wave spectra caused by spinor fields during inflation. These works have used the spinor mode sums in momentum space and have arrived at the fermionic field corrections to the gravitational wave spectra.
In this work, we aim to determine the coordinate space expression of the noise kernel for spinor fields in general FRW spacetimes. In order to carry out this task for the particular case of spinor fields in de Sitter spacetime, we consider that the fields are placed in the fermionic Bunch Davies vacuum \cite{Collins:2004wj,Koksma:2009tc} which is defined by choosing the fermionic mode functions in such a way that, in the asymptotic past, the leading order behaviour of these mode functions is like that of the Minkowski spacetime positive and negative frequency mode functions. Then, we deal with massless spinor fields in general FRW spacetimes and for these cases, we employ a mapping that relates a massless spinor field in any FRW spacetime to another massless spinor field in any other FRW spacetime and also relates the corresponding Wightman functions. Therefore, using this mapping, we place a massless spinor field in a given FRW spacetime in the Bunch-Davies vacuum of the corresponding massless spinor field in de Sitter spacetime (or equivalently, in terms of the Poincare vacuum of the Minkowski spacetime). We also compare the behaviour of the noise kernel of spinor fields in the above spacetimes with the analogous studies for scalar fields in the same spacetimes \cite{Dhanuka:2020yxp}. Since different epochs of the universe are approximately FRW type spacetimes, the analysis of massless spinor fields in FRW spacetimes is applicable to these different epochs of the universe, in particular, to the present day dark energy driven universe. As the surveys \cite{Planck:2018vyg} put the dark-energy equation of state parameter to be $-1.03 \pm 0.03$, having knowledge of the behaviour of the noise kernel for spinor fields in both the quintessence regime ($-1<w<-\frac{1}{3}$) and the phantom regime ($w <-1$) may be useful \cite{Johri:2003rh}. Therefore, our analysis considers spinor fields not only for quintessence but also for phantom cosmologies and in fact, for all types of FRW spacetimes. Carrying out this analysis for phantom cosmologies becomes physically more sound and, hence, all the more appealing when we can evade the unwanted big rip feature (i.e., the scaling factor and the energy density going to infinity in a finite interval of time) of the phantom models \cite{Johri:2003rh,Caldwell:1999ew} with a dynamical equation of state parameter generated by properly chosen phantom field potentials \cite{Singh:2003vx}. These types of potential allow for the possibility of having a phantom spacetime phase for some time and then exiting this phantom phase before big rip occurs. \\
The rest of the paper is organized as follows. In section $2$, we give a brief review of the dynamics of spinor fields in curved spacetimes and also provide an outline of a calculation to arrive at the expression of the Wightman function corresponding to the fermionic Bunch Davies vacuum of the de Sitter spacetime. We then employ a mapping between massless spinor fields in de Sitter and general FRW spacetimes and, using this mapping, we relate the expression of the Wightman function of a massless spinor field in de Sitter spacetime with the corresponding expressions for massless spinor fields in general FRW spacetimes. In section $3$, we use the point splitting technique to write the noise kernel for spinor fields as an expression involving some derivative operators acting on a product of Wightman functions. In section $4$, we arrive at the expression of the noise kernel for a spinor field (placed in the Bunch Davies vacuum) in de Sitter spacetime and analyse how the noise kernel behaves as a function of the mass of the field. In section $5$, we perform a similar analysis for massless fermions in general FRW spacetimes. In section $6$, we summarize our findings and discuss their implications. We use the $(-,+,+,+)$, convention for the spacetime metric.
\section{Preliminaries}
The dynamics of spinor fields in curved spacetimes is best described using the formalism of tetrad bases. The action for a minimally coupled spinor field in curved spacetime is given by
\begin{eqnarray}
\hspace{2.2cm} S &=& \int d^4x \sqrt{-g}\big[i\bar{\psi}\gamma^{\mu}\nabla_{\mu}\psi-m\bar{\psi}\psi\big] \nonumber \\ &=& \int d^4x \frac{\sqrt{-g}}{2}\big[i\bar{\psi}\gamma^{\mu}\nabla_{\mu}\psi - i(\nabla_{\mu}\bar{\psi})\gamma^{\mu}\psi -2m\bar{\psi}\psi\big] \nonumber \\&&  \hspace{2cm}+ \int d^4x \frac{\sqrt{-g}}{2}\nabla_{\mu}\big(i\bar{\psi}\gamma^{\mu}\psi\big)\, .
\end{eqnarray}
where $\gamma^{\mu} = e^{\mu}_{a}\Gamma^{a}$, $\{\Gamma^{a},\Gamma^{b}\} = -2\eta^{ab}$ and the four divergence term is a boundary term. Here $e^{\mu}_{a}$ are the tetrad basis which satisfy $e^{\mu}_{a}e^{\nu}_{b}\eta^{ab} = g^{\mu\nu}$. Also $\nabla_{\mu} = \partial_{\mu} - \frac{1}{8}\omega^{ab}_{\,\  \mu}[\Gamma_a,\Gamma_{b}]$ where $\omega^{ab}_{\,\ \mu} = e^{a}_{\lambda}e^{\tau b}\Gamma^{\lambda}_{\tau \mu} - e^{\tau b}\partial_{\mu}e^{a}_{\tau}$ (See \cite{Parker:2009uva,Shapiro:2016pfm}).\\ 
The equation of motion for the above action has the following form
\begin{equation}
(i\gamma^{\lambda}\nabla_{\lambda} - m)\psi = 0.
\end{equation}
Since we are ultimately interested in general FRW spacetimes for which one can choose coordinate systems in which the metric can be cast into conformally flat form \cite{Spradlin:2001pw}, let us consider $ds^2 = a^2(\eta)(-d\eta^2 + d\vec{x}^2)$. For these spacetimes, we can take the tetrad basis to be $e^{\mu}_{a} = \frac{1}{a} \delta^{\mu}_{a}$. Using this tetrad basis and the fact that $\Gamma^{\lambda}_{\mu\nu} = \frac{a'}{a}(\delta^{\lambda}_{\nu}\delta^{0}_{\mu} + \delta^{\lambda}_{\mu}\delta^{0}_{\nu} - \eta_{\mu\nu}\eta^{\lambda 0})$, we can calculate $\omega_{\mu a b} = \frac{a'}{a}(\eta_{a\mu}\delta^{0}_{b} - \eta_{b\mu}\delta^{0}_{a})$ where $'$ represents a derivative w.r.t. $\eta$. Using these expressions, we find that 
\begin{equation}
i\gamma^{\lambda}\nabla_{\lambda} = \frac{i}{a}\Gamma^{\mu}\bigg(\partial_{\mu} - \frac{1}{8}\omega_{  \mu ab}[\Gamma^a,\Gamma^{b}]\bigg) = \frac{i}{a}\bigg(\Gamma^{\mu}\partial_{\mu} + \frac{a'}{4a}\eta_{\mu a}\Gamma^{\mu}[\Gamma^0,\Gamma^{a}]\bigg) = \frac{i}{a}\bigg(\Gamma^{\mu}\partial_{\mu} + \frac{3a'}{2a}\Gamma^0\bigg) \, .
\end{equation}
Substituting this into the equation of motion, it modifies to
\begin{equation}
\bigg(i\Gamma^{\mu}\partial_{\mu} + \frac{3ia'}{2a}\Gamma^0 - a m \bigg)\psi = 0 \, .
\end{equation} 
Going to the Fourier space i.e., writing $\psi(\eta,\vec{x}) = \int \frac{d^3\vec{k}}{(2\pi)^3} e^{i\vec{k}.\vec{x}}\psi_{\vec{k}}(\eta)$, we obtain that 
\begin{equation}
\bigg(i\Gamma^{0}\partial_{0} -\vec{k}.\vec{\Gamma}+ \frac{3ia'}{2a}\Gamma^0 - a m \bigg)\psi_{\vec{k}}(\eta) = 0   \, ,
\end{equation}
and, for field redefinition $\chi_{\vec{k}}(\eta) = a^{\frac{3}{2}}(\eta)\psi_{\vec{k}}(\eta)$, it reduces further to the following form
\begin{equation}
\bigg(i\Gamma^{0}\partial_{0} -\vec{k}.\vec{\Gamma}- a m \bigg)\chi_{\vec{k}}(\eta) = 0  \, .
\end{equation}
Following \cite{Koksma:2009tc}, let us now decompose the 4-column spinors into chirality-helicity basis i.e.,  
\begin{gather}
\chi^{h}(\vec{k},\eta) \equiv \chi_{\vec{k}}(\eta) = \begin{bmatrix} \chi_{L,h}(\vec{k},\eta) \\ \chi_{R,h}(\vec{k},\eta)\end{bmatrix}\otimes \xi_{h}
 \end{gather}
where $\xi_{h}$ are 2-column eigenvectors of $(\hat{k}.\vec{\sigma})$ i.e., $(\hat{k}.\vec{\sigma})\xi_{h} = h\xi_{h}$ with eigenvalues $h$ being $\pm 1$\footnote{Using the form of the Pauli matrices and demanding that $\xi^{\dagger}_{h}\xi_{h} =1$, we can take the helicity eigenvectors to be
\begin{gather}
 \xi_{+} = \frac{1}{\sqrt{2(1-\hat{k}_{z})}}\begin{bmatrix} \hat{k}_{x} -i\hat{k}_{y}  \\ 1-\hat{k}_{z} \end{bmatrix},\,  \xi_{-} = \frac{1}{\sqrt{2(1+\hat{k}_{z})}}\begin{bmatrix} - \hat{k}_{x} + i\hat{k}_{y}  \\ 1+\hat{k}_{z} \end{bmatrix} \nonumber
 \, ,
\end{gather} 
where $\hat{\vec{k}} = \frac{\vec{k}}{||\vec{k}||}$.}. The functions $\chi_{L,h}(\vec{k},\eta)$ and $ \chi_{R,h}(\vec{k},\eta)$ are 1-column functions and are called left- and right-handed spinors of helicity $h$, respectively. Here $\otimes$ is the tensor product symbol.\\
Using the Weyl representation for Gamma matrices i.e.,  
\begin{gather}
 \Gamma^{0} = \begin{bmatrix} 0 & 1 \\ 1 & 0\end{bmatrix}\otimes \begin{bmatrix} 1 & 0 \\ 0 & 1\end{bmatrix}
 \, , \, \, \,  \Gamma^{i} = \begin{bmatrix}
   0 & 1\\
   -1 & 0
   \end{bmatrix} \otimes \sigma_{i}
\end{gather}
where $\otimes$ is, again, the tensor product symbol and $\sigma_i's$ are $2*2$ Pauli matrices, the Dirac equation reduces to the following set of coupled linear differential equations
 \begin{eqnarray}\label{2}
   i\partial_{0}\chi_{R,h}(\vec{k},\eta) - kh\chi_{R,h}(\vec{k},\eta)  - am\chi_{L,h}(\vec{k},\eta) &=& 0 \nonumber\\ i\partial_{0}\chi_{L,h}(\vec{k},\eta) + kh\chi_{L,h}(\vec{k},\eta)- am\chi_{R,h}(\vec{k},\eta) &=& 0 \, .
  \end{eqnarray}  
Defining $u_{\pm h}(k, \eta)  =\frac{\chi_{L,h}(\vec{k},\eta) \pm \chi_{R,h}(\vec{k},\eta)}{\sqrt{2}}$ and specializing to the de Sitter spacetime case i.e., taking $a(\eta) = -\frac{1}{H\eta}$, we get the following equations
\begin{equation}\label{1}
u''_{\pm h} + \Big(k^2 + \frac{\frac{1}{4}-(\frac{1}{2} \mp \frac{im}{H})^2}{\eta^2}\Big)u_{\pm h} =0 \, ,
\end{equation} 
where $'$ represents a derivative w.r.t. to $\eta$.
These equations have the form of Bessel's equation and hence the most general solutions for $u_{\pm h}$ are given by \cite{gradshteyn2014table,NIST:DLMF} 
\begin{equation}
u_{\pm h}(k,\eta) = \alpha^{h}_{\pm k}\sqrt{-k\eta} H^{(1)}_{\nu_{\pm}}(-k\eta) + \beta^{h}_{\pm k}\sqrt{-k\eta}H^{(2)}_{\nu_{\pm}}(-k\eta)
\end{equation}
where $\nu_{\pm} = \frac{1}{2}\mp \frac{im}{H}$ and $\alpha_{\pm k}^{h}$'s and $\beta_{\pm k}^{h}$'s are arbitrary constants to be fixed by initial condition. In order to proceed further with the analysis, we need to make a choice of vacuum for the spinor field and in this work, we choose to work with what is called the fermionic Bunch Davies vacuum.   
\subsection{The fermionic Bunch Davies vacuum}
The Bunch Davies vacuum for spinor fields in de Sitter spacetime is defined analogously to how one defines it for scalar fields. For this, we fix $\alpha_{\pm k}^{h}$'s and $\beta_{\pm k}^{h}$'s in such a way that the mode functions $u_{\pm h}(k,\eta)$, considered in the asymptotic past, behave like flat spacetime mode functions. To make it more precise, we notice that in the asymptotic past i.e., $\eta \to -\infty$ limit, equation (\ref{1}) reduces to $u''_{\pm h} + k^2 u_{\pm h} =0$. This implies that there exists positive and negative frequency modes in the asymptotic past i.e., $ e^{\pm ik\eta}$. Using the following large argument expansions of Hankel functions \cite{NIST:DLMF} i.e.,
\begin{eqnarray}
H^{(1)}_{\nu}(z) &= & \sqrt{\frac{2}{\pi z}}e^{i(z - \frac{\pi}{2}\nu - \frac{\pi}{4})}\Big[1 + O\Big(\frac{1}{z}\Big)\Big]  \, , \\
H^{(2)}_{\nu}(z) &= & \sqrt{\frac{2}{\pi z}}e^{-i(z - \frac{\pi}{2}\nu - \frac{\pi}{4})}\Big[1 + O\Big(\frac{1}{z}\Big)\Big] \, ,  \\  & & \hspace{2cm}\, \, \,  for \, \, |z| \to \infty, \,  \, \, Re(\nu) > -\frac{1}{2}\, \, \, and \, \, |arg(z)|<\pi, 
\end{eqnarray}
we see that, for non-zero $\alpha_{\pm k}^{h}$'s and vanishing $\beta_{\pm k}^{h}$'s, the mode functions, $u_{\pm h}(k,\eta)$, indeed behave like positive frequncy (or particle) modes in the $\eta \to -\infty$ limit. In fact, by substituting $u_{\pm h}(k,\eta) = \alpha^{h}_{\pm k}\sqrt{-k\eta} H^{(1)}_{\nu_{\pm}}(-k\eta)$ in equations of motion and using certain properties of Hankel functions, we find that 
$\alpha^{h}_{- k} = ih\alpha^{h}_{+ k} e^{i\pi \nu_{-}}.$
Demanding that $\lim_{\eta \to -\infty}u_{+ h}(k,\eta) = \frac{e^{-ik\eta}}{\sqrt{2}}$, we should take $\alpha^{h}_{+ k} = \sqrt{\frac{\pi}{4}}e^{i\frac{\pi}{2}(\nu_{+} + 1/2)}$ and therefore, we have, for particle modes, 
\begin{eqnarray}
u_{+ h}(k,\eta) &=& \sqrt{\frac{-\pi k\eta}{4}}e^{i\frac{\pi}{2}(\nu_{+} + 1/2)} H^{(1)}_{\nu_{+}}(-k\eta) \equiv f(k,\eta)\\
u_{- h}(k,\eta) &=& -h\sqrt{\frac{-\pi k\eta}{4}}e^{i\frac{\pi}{2}(\nu_{-} + 1/2)} H^{(1)}_{\nu_{-}}(-k\eta)  \equiv -h g^{*}(k,\eta)\, .
\end{eqnarray}
This implies that 
\begin{eqnarray}
\chi_{L,h}(k,\eta) &=& \frac{f - h g^{*}}{\sqrt{2}}\\
\chi_{R,h}(k,\eta) &=& \frac{f + h g^{*}}{\sqrt{2}} \, .
\end{eqnarray}
Similarly, we can determine the  anti-particle modes. For anti-particle modes, we denote the Fourier coefficient of $e^{-i\vec{k}.\vec{x}}$ to be $\nu_{\vec{k}}(\eta)$ and hence, in the helicity-chirality basis, the differential equations have $-k$ in place of $k$. Therefore, by flipping the definitions of the left and right handed fermions i.e., taking 
\begin{gather}
 \nu^{h}(\vec{k},\eta) \equiv \nu_{\vec{k}}(\eta) = \begin{bmatrix} \nu_{R,h}(\vec{k},\eta) \\ \nu_{L,h}(\vec{k},\eta)\end{bmatrix}\otimes \xi_{h} \, ,
 \end{gather}
 we end up with the same set of differential equations for $\nu_{R,h}$ and $\nu_{L,h}$ as (\ref{2}). Now demanding that the corresponding $u_{\pm h}$'s go to $\frac{e^{ik\eta}}{\sqrt{2}}$ in the asymptotic past, we conclude that 
 \begin{eqnarray}
\nu_{L,h}(k,\eta) &=& \frac{h f^{*} + g}{\sqrt{2}}\\
\nu_{R,h}(k,\eta) &=& \frac{- h f^{*} + g}{\sqrt{2}} \, .
\end{eqnarray}
\\ Using the above considered Bunch Davies modes, we can write the field operator and its conjugate as follows\footnote{For the rest of this subsection, $a(\eta)$ represents $-\frac{1}{H\eta}$.}
\begin{equation}
\hat{\psi} (\eta,\vec{x}) = a^{-\frac{3}{2}}(\eta)\int \frac{d^3\vec{k}}{(2\pi )^3}\sum_{h}\Big[ \hat{a}_{\vec{k},h} \chi^{h}(\vec{k},\eta)e^{i\vec{k}.\vec{x}}  + \hat{b}^{\dagger}_{\vec{k},h}\nu^{h}(\vec{k},\eta) e^{-i\vec{k}.\vec{x}} \Big]
\end{equation}
and
\begin{equation}
\hat{\bar{\psi}} (\eta,\vec{x}) = a^{-\frac{3}{2}}(\eta)\int \frac{d^3\vec{k}}{(2\pi )^3}\sum_{h}\Big[ \hat{a}^{\dagger}_{\vec{k},h} \bar{\chi}^{h}(\vec{k},\eta)e^{-i\vec{k}.\vec{x}}  + \hat{b}_{\vec{k},h}\bar{\nu}^{h}(\vec{k},\eta) e^{i\vec{k}.\vec{x}} \Big] \, .
\end{equation}
Here $\hat{a}$'s and $\hat{b}$'s are the annihilation operators while  $\hat{a}^{\dagger}$'s and $\hat{b}^{\dagger}$'s are the creation operators corresponding to the above described Bunch Davies modes. We now define the Bunch Davies vacuum, $\ket{0}$, to be the state which is annihilated by all $\hat{a}_{\vec{k},h}$'s and $\hat{b}_{\vec{k},h}$'s i.e., $\hat{a}_{\vec{k},h}\ket{0} = 0$ and $\hat{b}_{\vec{k},h}\ket{0} = 0$.  \\
From the expressions for the field operator and its adjoint, we find that the Wightman functions are given by 
\begin{equation}
S_{ij}(x,x') = \langle\hat{\psi}_{i}(x)\hat{\bar{\psi}}_{j}(x')\rangle = a^{-\frac{3}{2}}(\eta)a^{-\frac{3}{2}}(\eta')\int \frac{d^3\vec{k}}{(2\pi)^3}\sum_{h}\chi^{h}_{i}(\vec{k},\eta)\bar{\chi}^{h}_{j}(\vec{k},\eta')e^{i\vec{k}.(\vec{x}-\vec{x}')}
\end{equation}
and 
\begin{multline}
R_{ji}(x',x) = \langle\hat{\bar{\psi}}_{j}(x')\hat{\psi}_{i}(x)\rangle = a^{-\frac{3}{2}}(\eta)a^{-\frac{3}{2}}(\eta')\int \frac{d^3\vec{k}}{(2\pi)^3}\sum_{h}\nu^{h}_{i}(-\vec{k},\eta)\bar{\nu}^{h}_{j}(-\vec{k},\eta')e^{i\vec{k}.(\vec{x}-\vec{x}')} \\ = -S_{ij}(x,x') \hspace*{8cm}
\end{multline}
After some manipulations, one can show that (See \cite{Koksma:2009tc, Collins:2004wj})
\begin{multline}
S_{ij}(x,x') = \langle\hat{\psi}_{i}(x)\hat{\bar{\psi}}_{j}(x')\rangle = a(\eta_{x})\Big[i\gamma^{\lambda}\overrightarrow{\nabla}^{x}_{\lambda} + m\Big]\frac{H^2}{\sqrt{a(\eta_{x})a(\eta_{x'})}}\Bigg[S_{+}(x,x')\frac{1 + \Gamma^{0}}{2} + S_{-}(x,x')\frac{1 - \Gamma^{0}}{2}\Bigg]
\end{multline}
where 
\begin{equation}
S_{\pm}(x,x') = \frac{\Gamma\big(2 \pm i\frac{m}{H}\big)\Gamma\big(1 \mp i\frac{m}{H}\big)}{(4\pi)^2}  {}_{2}F_1\Big(2 \pm i\frac{m}{H}, 1 \mp i\frac{m}{H},2,Z(x,x')\Big)
\end{equation}
and $Z(x,x') = 1 + \frac{(\eta-\eta')^2 - (\Delta \vec{x})^2}{4\eta\eta'}$. Using that $\overrightarrow{\nabla}^{x}_{\lambda} = \partial_{\lambda} - \frac{1}{8}\omega_{ab\lambda}[\Gamma^a,\Gamma^b] = \partial_{\lambda} +\frac{a'}{4a}\eta_{\lambda a}[\Gamma^{0},\Gamma^a]$ and $\gamma^{\lambda}\overrightarrow{\nabla}^{x}_{\lambda} = \frac{1}{a}\big(\Gamma^{\lambda}\partial_{\lambda} + \frac{3a'}{2a}\Gamma^{0}\big)$, the above expression can be written as 
\begin{multline}
S_{ij}(x,x') = \Big[i\Gamma^{\lambda}\partial^{x}_{\lambda} + \frac{3i a'(\eta_x)}{2a(\eta_x)}\Gamma^0+ a(\eta_x)m\Big]\frac{H^2}{\sqrt{a(\eta_{x})a(\eta_{x'})}} \Bigg[S_{+}(x,x')\frac{1 + \Gamma^{0}}{2} + S_{-}(x,x')\frac{1 - \Gamma^{0}}{2}\Bigg] \\   =  \frac{H^2}{\sqrt{a(\eta_{x})a(\eta_{x'})}}\Big[i\Gamma^{\lambda}\partial^{x}_{\lambda} + i\frac{ a'(\eta_x)}{a(\eta_x)}\Gamma^0+ a(\eta_x)m\Big] \Bigg[S_{+}(x,x')\frac{1 + \Gamma^{0}}{2} + S_{-}(x,x')\frac{1 - \Gamma^{0}}{2}\Bigg]  \, .
\end{multline}
We need the forms of the Wightman functions, $S_{ij}(x,x')$ and $R_{ji}(x',x)$, later on in this work when we study the behaviour of the fluctuations in the stress energy operator which, as we shall see, can be expressed as certain derivatives acting on products of Wightman functions.  
\subsection{Equivalence between massless spinor fields in FRW spacetimes and de Sitter spacetime}
It has been established, in \cite{Lochan:2018pzs}, that massless scalar fields in FRW spacetimes can be mapped to massive scalar fields in de Sitter spacetime and hence one can make use of this equivalence to study the dynamics in one setup in terms of the another. In this subsection, we make use of the conformal invariance of massless spinor fields and relate the Wightman functions of massless spinor fields in general FRW spacetimes with the Wightman function of a massless spinor field in de Sitter spacetime. In fact, we start with a massive spinor field in an arbitrary FRW spacetime and try to relate it to another spinor field in some other arbitrary FRW spacetime. For this, consider the action of a spinor field in a spacetime with $ds^2 = c^{2}(\eta)(-d\eta^2 + d\vec{x}^2)$ i.e., 
\begin{eqnarray}
S &=& \int d^4x \sqrt{-g}\big[i\bar{\psi}\gamma^{\mu}\nabla_{\mu}\psi-m\bar{\psi}\psi\big] \nonumber \\
& = & \int d^4x c^3\bar{\psi}\Big[i\Gamma^{\mu}\partial_{\mu} + i\frac{3c'}{2c}\Gamma^{0}-c m\Big]\psi \, .
\end{eqnarray}
Considering the transformation $\psi = F(\eta)\Omega$, we observe that the above action transforms to 
\begin{eqnarray}
S &=& \int d^4x c^3F^2\bar{\Omega}\Big[i\Gamma^{\mu}\partial_{\mu} + i\frac{3c'}{2c}\Gamma^{0} + i\frac{F'}{F}\Gamma^{0} -c m\Big]\Omega
\end{eqnarray}
and now demanding that $c^3F^2 = b^3$ and $\Big[i\frac{3c'}{2c}\Gamma^{0} + i\frac{F'}{F}\Gamma^{0} -c m\Big] = \Big[i\frac{3b'}{2b}\Gamma^{0} - b m'\Big] $, the above action becomes
\begin{eqnarray}
S &=& \int d^4x b^3\bar{\Omega}\Big[i\Gamma^{\mu}\partial_{\mu} + i\frac{3b'}{2b}\Gamma^{0} - b m'\Big]\Omega  
\end{eqnarray}
which is the action of a spinor field, $\Omega$, in a spacetime with $ds^2 = b^{2}(\eta)(-d\eta^2 + d\vec{x}^2)$. 

\noindent We notice that the condition that $c^3F^2 = b^3$ automatically gives
\begin{equation}  \frac{3c'}{2c} + \frac{F'}{F} = \frac{3b'}{2b} \, .
\end{equation}
Therefore, from the condition that 
\begin{equation}
\Big[i\frac{3c'}{2c}\Gamma^{0} + i\frac{F'}{F}\Gamma^{0} -c m\Big] = \Big[i\frac{3b'}{2b}\Gamma^{0} - b m'\Big] \, ,
\end{equation} 
we obtain that $m' = \frac{c}{b}m$. A number of interesting conclusions can be drawn from this analysis. But for our purposes, all we need to observe is that a massless spinor field in an FRW spacetime can always be mapped to another massless spinor field in any other FRW spacetime, in particular we can map a massless spinor field in any FRW spacetime to a massless spinor field in de Sitter or flat spacetime.
We make use of this equivalence and determine the desired results in general FRW spacetimes in terms of the de Sitter quantities\footnote{As we have seen that massless spinor fields in FRW spacetimes are equivalent to massless spinor fields in flat spacetime as well, we could have equally well carried out the FRW analyses in terms of the flat space-time quantities. But since we are going to analyse the de Sitter noise kernel before going to the general FRW cases, it seems more seamless with the flow of the paper to map FRW analyses to the de Sitter case.} i.e., we take $b(\eta) = a(\eta) = -\frac{1}{H\eta}$. In particular, we observe that the Wightmann function for a massless spinor field in an FRW spacetime (with scale factor being $c(\eta)$) is given by
\begin{multline}
S^{FRW}_{ij}(x,x') = \langle\psi_{i}(x)\bar{\psi}_{j}(x')\rangle = (F(\eta)F(\eta'))\langle\Omega_{i}(x)\bar{\Omega}_{j}(x')\rangle = \Big(\frac{a(\eta)}{c(\eta)}\Big)^{\frac{3}{2}}\Big(\frac{a(\eta')}{c(\eta')}\Big)^{\frac{3}{2}}S^{dS}_{ij}(x,x') \\ 
 = \frac{1}{{c^{\frac{3}{2}}(\eta_{x})c^{\frac{3}{2}}(\eta_{x'})}}S^{flat}_{ij}(x,x') \, .
\end{multline}
Similarly, we have 
\begin{multline}
R^{FRW}_{ji}(x',x) = \langle\bar{\psi}_{j}(x')\psi_{i}(x)\rangle = (F(\eta)F(\eta'))\langle\bar{\Omega}_{j}(x')\Omega_{i}(x)\rangle = \Big(\frac{a(\eta)}{c(\eta)}\Big)^{\frac{3}{2}}\Big(\frac{a(\eta')}{c(\eta')}\Big)^{\frac{3}{2}}R^{dS}_{ji}(x',x)  \\
=  \frac{1}{{c^{\frac{3}{2}}(\eta_{x})c^{\frac{3}{2}}(\eta_{x'})}}R^{flat}_{ji}(x',x)\, .
\end{multline}
We make use of these relations later on in this work and find that the noise kernel for massless spinor fields in general FRW spacetimes is related to the de Sitter noise kernel again by conformal factors. 
\section{Noise kernel and point splitting} 
Using the expression that
\begin{equation}
T_{\mu\nu}(x) = -\frac{2}{\sqrt{-g}}\frac{\delta S}{\delta g^{\mu\nu}(x)} \, ,
\end{equation} 
we find that, for spinor fields in curved spacetimes, the stress energy tensor \cite{Parker:2009uva,Shapiro:2016pfm} is given by
\begin{eqnarray}
T_{\mu\nu} &=& -\frac{i}{2}g_{\mu\nu}\big[\bar{\psi}\gamma^{\lambda}\overrightarrow{\nabla}_{\lambda}\psi-\bar{\psi}\overleftarrow{\nabla}_{\lambda}\gamma^{\lambda}\psi\big] + \frac{i}{2}\big[\bar{\psi}\gamma_{(\mu}\overrightarrow{\nabla}_{\nu)}\psi-\bar{\psi}\overleftarrow{\nabla}_{(\nu}\gamma_{\mu)}\psi\big]  + m\bar{\psi}\psi g_{\mu\nu} \nonumber \\
&=& -\frac{g_{\mu\nu}}{2}\bar{\psi}\Big[(i\gamma^{\lambda}\overrightarrow{\nabla}_{\lambda}- m) - (i\overleftarrow{\nabla}_{\lambda}\gamma^{\lambda} + m )\Big]\psi + \frac{i}{2}\bar{\psi}\big[\gamma_{(\mu}\overrightarrow{\nabla}_{\nu)}-\overleftarrow{\nabla}_{(\nu}\gamma_{\mu)}\big]\psi \, .
\end{eqnarray}
Using the point splitting technique \cite{Bunch:1978yq}, this can succinctly be written as 
\begin{equation}
T_{\mu\nu}(x) = \lim_{x' \to x}P_{\mu\nu ij}(x,x')\bar{\psi}_i(x) \psi_j(x') \, ,
\end{equation}
where summation over $i$ and $j$ is understood and 
\begin{multline}
P_{\mu\nu ij}(x,x') 
= -\frac{g_{\mu\nu}}{2}\Big[(i\gamma^{\lambda}\overrightarrow{\nabla}^{x'}_{\lambda}- m) - (i\overleftarrow{\nabla}^{x}_{\lambda}\gamma^{\lambda} + m )\Big]_{ij} + \frac{i}{2}\big[\gamma_{(\mu}\overrightarrow{\nabla}^{x'}_{\nu)}-\overleftarrow{\nabla}^{x}_{(\nu}\gamma_{\mu)}\big]_{ij} \, .
\end{multline}
To conclude whether the semiclassical analyses (based only on the expectation values of the stress energy operators) alone are sufficient or not, we must have knowledge of the behaviour of the quantum fluctuations in stress energy tensors. In stochastic gravity paradigm \cite{Hu:2008rga}, these are characterised by the noise kernel which is given by
\begin{equation}
N_{abcd}(x,y) = \langle \hat{T}_{ab}(x)\hat{T}_{cd}(y)\rangle - \langle \hat{T}_{ab}(x)\rangle \langle \hat{T}_{cd}(y)\rangle \, ,
\end{equation}
where, as usual, the quantum stress energy operator is obtained from the classical expression by replacing the classical fields by their corresponding operator quantities.\\
Using the above notations, we find that the noise kernel can be written as 
\begin{multline}
N_{abcd}(x,y) = \lim_{x' \to x}\lim_{y' \to y}P_{ab ij}(x,x')P_{cd kl}(y,y')\Big[\langle \hat{\bar{\psi}}_i(x) \hat{\psi}_j(x')\hat{\bar{\psi}}_k(y) \hat{\psi}_l(y')\rangle\\ - \langle \hat{\bar{\psi}}_i(x) \hat{\psi}_j(x')\rangle \langle \hat{\bar{\psi}}_k(y) \hat{\psi}_l(y')\rangle\Big] \, .
\end{multline}
Applying Wick's theorem to the first term, we see that the noise kernel is now given by 
\begin{equation}
N_{abcd}(x,y) = \lim_{x' \to x}\lim_{y' \to y}P_{ab ij}(x,x')P_{cd kl}(y,y')\Big[\langle \hat{\bar{\psi}}_i(x) \hat{\psi}_l(y')\rangle \langle \hat{\psi}_j(x')\hat{\bar{\psi}}_k(y) \rangle\Big] \, .
\end{equation}
Since $\hat{\psi}$ is a linear combination of the solutions of the Dirac equation and $\hat{\bar{\psi}}$ is a linear combination of the solutions of the adjoint of the Dirac equation, we see that, in the stress energy operator, the terms which correspond to the Dirac equation and its adjoint drop out. Therefore, we are left with the following expression for the noise kernel
\begin{multline}\label{3}
N_{abcd}(x,y) = \lim_{x' \to x}\lim_{y' \to y}\frac{1}{4}\big[\gamma_{(a}\overrightarrow{\nabla}^{x'}_{b)}-\overleftarrow{\nabla}^{x}_{(a}\gamma_{b)}\big]_{ij}\big[\gamma_{(c}\overrightarrow{\nabla}^{y'}_{d)}-\overleftarrow{\nabla}^{y}_{(c}\gamma_{d)}\big]_{kl}S_{jk}(x',y)S_{li}(y',x) \\ =  \lim_{x' \to x}\lim_{y' \to y}\frac{1}{4}\Big[Tr\big(\gamma_{(a}\overrightarrow{\nabla}^{x'}_{b)}S(x',y)\gamma_{(c}\overrightarrow{\nabla}^{y'}_{d)}S(y',x)\big) - Tr\big(\gamma_{(a}\overrightarrow{\nabla}^{x'}_{b)}S(x',y)\overleftarrow{\nabla}^{y}_{(c}\gamma_{d)}S(y',x)\big)\\ -Tr\big(S(x',y)\gamma_{(c}\overrightarrow{\nabla}^{y'}_{d)}S(y',x)\overleftarrow{\nabla}^{x}_{(a}\gamma_{b)}\big) + Tr\big(S(x',y)\overleftarrow{\nabla}^{y}_{(c}\gamma_{d)}S(y',x)\overleftarrow{\nabla}^{x}_{(a}\gamma_{b)}\big) \Big]  \, .
\end{multline}
The analysis done in this section is independent of the spacetime that we consider and as such holds for all types of spacetimes. In the next section, we calculate the behaviour of the noise kernel in de Sitter spacetime while placing the quantum spinor field in the Bunch Davies vacuum and therefore we use the Wightmann function for this state given in the previous section. In particular, we calculate the $(a=b=c=d=0)$ component of the noise kernel as this can be related to an invariant quantity called the energy-energy correlator as discussed later on in this work. 

\section{Noise kernel in de Sitter spacetime}
In this section, we specialize to the case of a spinor field evolving in the de Sitter spacetime. As mentioned in the previous section, we are interested in calculating the $(a=b=c=d=0)$ component of the noise kernel i.e., $N_{0000}(x,y)$. Since we have expressed the noise kernel as a sum of derivatives acting on a product of Wightman functions, we take the above given expression of the Wightman function for a spinor field (in de Sitter spacetime) placing it in the Bunch Davies vacuum. Using that $\overrightarrow{\nabla}^{x}_{0} = \partial^{x}_{0}$, we see that\footnote{From this section onwards, we always take $a_{x} \equiv a(\eta_{x}) = -\frac{1}{H\eta_{x}} $}
\begin{multline}\label{4}
N_{0000}(x,y) =  \lim_{x' \to x}\lim_{y' \to y}\frac{a(\eta_x)a(\eta_y)}{4}\Bigg[\partial^{x'}_{0}\partial^{y'}_{0} + \partial^{x}_{0}\partial^{y}_{0} - \partial^{x'}_{0}\partial^{y}_{0} - \partial^{x}_{0}\partial^{y'}_{0}\Bigg]Tr\big(\Gamma_{0}S(x',y)\Gamma_{0}S(y',x)\big)\\  =  \lim_{x' \to x}\lim_{y' \to y}\frac{a_{x}a_{y}}{4}\Bigg[\partial^{x'}_{0}\partial^{y'}_{0} + \partial^{x}_{0}\partial^{y}_{0} - \partial^{x'}_{0}\partial^{y}_{0} - \partial^{x}_{0}\partial^{y'}_{0}\Bigg]\frac{H^4}{\sqrt{a_{x}a_{x'}a_{y}a_{y'}}} \\ Tr\Bigg(\Gamma_{0}\Big[i\Gamma^{\lambda}\partial^{x'}_{\lambda} + i\frac{ a'_{x'}}{a_{x'}}\Gamma^0+ a_{x'}m\Big] \Big[\sum_{\epsilon=\pm}S_{\epsilon}(x',y)\frac{1 + \epsilon\Gamma^{0}}{2} \Big]\\ \Gamma_{0}\Big[i\Gamma^{\sigma}\partial^{y'}_{\sigma} + i\frac{ a'_{y'}}{a_{y'}}\Gamma^0+ a_{y'}m\Big] \Big[\sum_{\epsilon=\pm}S_{\epsilon}(y',x)\frac{1 + \epsilon\Gamma^{0}}{2} \Big]\Bigg) \, .
\end{multline}
It can be easily shown that the factor $\frac{H^4}{\sqrt{a_{x}a_{x'}a_{y}a_{y'}}}$, after having been operated by all the derivative operators, simply come in front of the bigger square brackets (which contain the derivative operators) as it is and in the above limits cancel out the already present $a_{x}a_{y}$ factor. Finally, we obtain 
\begin{multline}\label{5}
N_{0000}(x,y) =  \lim_{x' \to x}\lim_{y' \to y}\frac{H^4}{4}\Bigg[\partial^{x'}_{0}\partial^{y'}_{0} + \partial^{x}_{0}\partial^{y}_{0} - \partial^{x'}_{0}\partial^{y}_{0} - \partial^{x}_{0}\partial^{y'}_{0}\Bigg] \\ Tr\Bigg(\Gamma_{0}\Big[i\Gamma^{\lambda}\partial^{x'}_{\lambda} + i\frac{ a'_{x'}}{a_{x'}}\Gamma^0+ a_{x'}m\Big] \Big[\sum_{\epsilon=\pm}S_{\epsilon}(x',y)\frac{1 + \epsilon\Gamma^{0}}{2} \Big]\\ \Gamma_{0}\Big[i\Gamma^{\sigma}\partial^{y'}_{\sigma} + i\frac{ a'_{y'}}{a_{y'}}\Gamma^0+ a_{y'}m\Big] \Big[\sum_{\epsilon=\pm}S_{\epsilon}(y',x)\frac{1 + \epsilon\Gamma^{0}}{2} \Big]\Bigg) \, .
\end{multline}
Thus, we see that, in the expression of the noise kernel, we have traces of the Dirac matrices. Therefore, we can use the well known properties \cite{Parker:2009uva} of these traces to proceed further with our calculations. 
\subsection{``Gammatronics"}
Using the cyclic property of traces, we can write the noise kernel expression in the following expanded form
\begin{multline}
N_{0000}(x,y) =  \lim_{x' \to x}\lim_{y' \to y}\frac{H^4}{4}\Bigg[\partial^{x'}_{0}\partial^{y'}_{0} + \partial^{x}_{0}\partial^{y}_{0} - \partial^{x'}_{0}\partial^{y}_{0} - \partial^{x}_{0}\partial^{y'}_{0}\Bigg] \\ \sum_{\epsilon=\pm}\sum_{\delta=\pm}Tr\Bigg(\Gamma_{0}\Big[i\Gamma^{\lambda}\partial^{x'}_{\lambda} + i\frac{ a'_{x'}}{a_{x'}}\Gamma^0+ a_{x'}m\Big] \Big[\frac{1 + \epsilon\Gamma^{0}}{2} \Big]\\ \Gamma_{0}\Big[i\Gamma^{\sigma}\partial^{y'}_{\sigma} + i\frac{ a'_{y'}}{a_{y'}}\Gamma^0+ a_{y'}m\Big] \Big[\frac{1 + \delta\Gamma^{0}}{2} \Big]\Bigg)S_{\epsilon}(x',y)S_{\delta}(y',x) \\ =
\lim_{x' \to x}\lim_{y' \to y}\frac{H^4}{4}\Bigg[\partial^{x'}_{0}\partial^{y'}_{0} + \partial^{x}_{0}\partial^{y}_{0} - \partial^{x'}_{0}\partial^{y}_{0} - \partial^{x}_{0}\partial^{y'}_{0}\Bigg]  \sum_{\epsilon=\pm}\sum_{\delta=\pm}Tr\Bigg(i^2\Gamma^{\lambda}M\Gamma^{\sigma}P\partial_{\lambda}^{x'}\partial_{\sigma}^{y'} \\ +  i^2 \Gamma^{\lambda}M\Gamma^{0}P\frac{a'_{y'}}{a_{y'}}\partial^{x'}_{\lambda} + ia_{y'}m \Gamma^{\lambda}MP\partial^{x'}_{\lambda} + i^2\frac{a'_{x'}}{a_{x'}}\Gamma^{0}M\Gamma^{\sigma}P\partial^{y'}_{\sigma} + i^2\frac{a'_{x'}}{a_{x'}}\frac{a'_{y'}}{a_{y'}}\Gamma^{0}M\Gamma^{0}P \\ + im\frac{a'_{x'}}{a_{x'}}a_{y'}\Gamma^{0}MP + ia_{x'}m M\Gamma^{\sigma}P\partial^{y'}_{\sigma}+ ia_{x'}m\frac{a'_{y'}}{a_{y'}} M\Gamma^{0}P + a_{x'}a_{y'}m^2MP\Bigg)S_{\epsilon}(x',y)S_{\delta}(y',x) 
\end{multline}
where $M = \Big[\frac{\Gamma^{0} + \epsilon}{2}\Big]$ and $P = \Big[\frac{\Gamma^{0} + \delta}{2}\Big]$. \\
Employing the known properties of the traces of the Gamma matrices, we find that the above expression reduces to 
\begin{multline}
N_{0000}(x,y) =
-\lim_{x' \to x}\lim_{y' \to y}\frac{2 H^4}{4}\Bigg[\partial^{x'}_{0}\partial^{y'}_{0} + \partial^{x}_{0}\partial^{y}_{0} - \partial^{x'}_{0}\partial^{y}_{0} - \partial^{x}_{0}\partial^{y'}_{0}\Bigg]  \\ \Bigg[\sum_{\epsilon = \pm}\Bigg(\partial_{0}^{x'}\partial_{0}^{y'}+ \Big(\frac{a'_{x'}}{a_{x'}} - \epsilon ima_{x'}\Big)\partial^{y'}_{0} + \Big(\frac{a'_{y'}}{a_{y'}} -\epsilon ima_{y'}\Big)\partial^{x'}_{0} + \Big(\frac{a'_{y'}}{a_{y'}} -\epsilon ima_{y'}\Big)\Big(\frac{a'_{x'}}{a_{x'}} -\epsilon ima_{x'}\Big)\Bigg)S_{\epsilon}(x',y)S_{\epsilon}(y',x) \\ + \delta^{kl}\partial^{x'}_{k}\partial^{y'}_{l}\Big(S_{+}(x',y)S_{-}(y',x) + S_{-}(x',y)S_{+}(y',x)\Big)\Bigg]\, .
\end{multline}
\subsection{Behaviour on equal time sheets}
In order to study how the energy densities for spatially separated points are correlated with each other, we specialize to the case where $\eta_x = \eta_y$. But before we consider this case, we have to perform all the derivatives above. While performing these calculations, we make use of the fact that $S_{\pm}(x,x')= S_{\pm}(Z(x,x'))$ i.e., it is a function of the invariant distance $Z$. We should take extra care to make sure that we take all the limits only after performing all the derivatives. Using the formulae given in Appendix A, we find that, on equal time sheets, we have 
\begin{multline}\label{10}
-\frac{2}{H^4}N_{0000}(x,y) = \big(S''_{+}S'_{-} + (- \leftrightarrow +)\big)\Big[-\frac{(\Delta\vec{x})^4}{8\eta^8}-\frac{(\Delta\vec{x})^2}{4\eta^6}\Big] + \big(S''_{+}S''_{-} )\Big[-\frac{(\Delta\vec{x})^6}{16\eta^{10}}\Big] \\ + \big(S'''_{+}S'_{-} + (- \leftrightarrow +)\big)\Big[\frac{(\Delta\vec{x})^6}{32\eta^{10}}\Big]  +
\sum_{\epsilon = \pm}\Big(1-\frac{\epsilon im}{H}\Big)\Bigg[\frac{(S'_{\epsilon})^2}{4}\Big[\frac{(\Delta \vec{x})^4}{2\eta^8}-\frac{(\Delta \vec{x})^2}{\eta^6}\Big] + \frac{2}{\eta^4}S_{\epsilon}S'_{\epsilon}+       S_{\epsilon}S''_{\epsilon}\Big[-\frac{3(\Delta \vec{x})^4}{8\eta^8}-\frac{(\Delta \vec{x})^2}{4\eta^6}\Big] \\ + (S_{\epsilon}S'''_{\epsilon}-S'_{\epsilon}S''_{\epsilon})\frac{(\Delta \vec{x})^6}{32\eta^{10}}\Bigg]  +
\Big(1-\frac{\epsilon im}{H}\Big)^2\Bigg[\frac{(S'_{\epsilon})^2}{4}\Big[\frac{(\Delta \vec{x})^4}{2\eta^8}\Big] + \frac{S_{\epsilon}S_{\epsilon}}{\eta^4}+       S_{\epsilon}S'_{\epsilon}\Big[\frac{(\Delta \vec{x})^2}{2\eta^6}+\frac{1}{\eta^4}\Big]  -S_{\epsilon}S''_{\epsilon}\frac{(\Delta \vec{x})^4}{8\eta^{8}}\Bigg]
\\ +
\Bigg[\frac{(S'_{\epsilon})^2}{4}\Big[\frac{(\Delta \vec{x})^4}{4\eta^8}-2\frac{(\Delta \vec{x})^2}{\eta^6}+ \frac{2}{\eta^4}\Big] + \frac{S'_{\epsilon}S''_{\epsilon}}{8}\frac{(\Delta \vec{x})^4}{4\eta^6}\Big[\frac{2}{\eta^2} + \frac{(\Delta \vec{x})^2}{\eta^4}\Big] +  \frac{S''_{\epsilon}S''_{\epsilon}}{16}\frac{(\Delta \vec{x})^8}{8\eta^{12}} - \frac{S'_{\epsilon}S'''_{\epsilon}}{16}\frac{(\Delta \vec{x})^8}{8\eta^{12}}     \Bigg] \, .
\end{multline}
Now we are interested in seeing the behaviour of this noise kernel for late time limits i.e., $\eta \to 0$ limit. Recalling that the $S_{\epsilon}'s$ are just Hypergeometric functions (see section $2$) and using the fact that their derivatives are again Hypergeometric functions \cite{NIST:DLMF} and also making use of their asymptotic behaviour, we find that the leading order behaviour of the noise kernel is given by
\begin{equation}
N_{0000}(x,y) = \frac{2H^4\eta^2}{\pi^3(\Delta \vec{x})^6}\frac{(1 + \frac{m^2}{H^2})(\frac{m^3}{H^3})}{\mathrm{sinh}(\frac{2\pi m}{H})} + O (\eta^4).
\end{equation}
Thus, from the above expression, we see that, irrespective of how massive or light the fermionic fields are, the considered component of the noise kernel vanishes in the late time limit i.e., $\eta \to 0 $ limit. In that sense, the decay of the noise kernel component for a spinor field in de Sitter spacetime is a universal phenomenon independent of the mass of the field. This is in contrast to the behaviour of the noise kernel for quantum scalar fields in de Sitter spacetime where the considered noise kernel component shows a transition from vanishing to divergent behaviour as mass is varied from $[0,\frac{3}{2}]$ with `critical mass' value being $\frac{m^2}{H^2} = 2$ (Refer \cite{Dhanuka:2020yxp}). This implies that, if we start with two spatially separated points on a constant time sheet in the past, then the corresponding local stress energy tensors (physically speaking, energy momentum content) don't develop any correlations in the far future. In fact, more appropriately, the correlations between the quantum stress energy tensor at different points get washed away as the de Sitter spacetime evolves to late time limit. To present this conclusion in a more illuminating manner, we notice that at the start of the inflation i.e., in the $\eta \to -\infty$ limit, all points, on this very early constant time slice of the de Sitter spacetime, are separated by zero physical distances and hence, the quantum field at different coordinate distances (but with zero physical separations) have maximum correlations. However, as the de Sitter spacetime evolves, the physical distances between spatially separated points increase and intuitively, we expect that the correlations between the corresponding local stress energy tensors (for that matter, between the field operators themselves or any other functions of the field operators) should decrease.  
The above obtained results suggest that the quantum dynamics of the spinor fields in de Sitter spacetime is not so that it can overcome the diminishing effect of increasing physical distances and we have an overall vanishing of the noise kernel in the late time limit. But we see (from the behaviour of the noise kernel of scalar fields \cite{Dhanuka:2020yxp}) that, for quantum scalar fields in de Sitter spacetime, there are mass ranges for which their dyanmics can overcome the effect of the classical expansion of the de Sitter spacetime and we obtain divergent noise kernel for these cases. Another important yet very intuitive property of the noise kernel is that the vanishing of the noise kernel is inversely proportional to the coordinate distances between spatially separated points i.e., the lesser the coordinate distances are (between the spatially separated points), the later (in the evolution history of the de Sitter spacetime) the correlations decay. \\ 
As the noise kernel is a (bi-)tensor quantity and hence is expressed differently in different coordinate systems, let us express the above obtained results in terms of a coordinate invariant quantity i.e., the energy-energy correlator (Refer \cite{Dhanuka:2020yxp}). To define this quantity, we consider a comoving observer (i.e., one whose spatial coordinates are fixed) for which the normalized tangent vector field along its trajectory is given by $t^{\alpha}(\eta,\vec{x}) = (\frac{1}{a(\eta)},0,0,0)$ (and hence $g_{\alpha\beta}t^{\alpha}t^{\beta}=-1$). One then defines the energy density at any point by the contraction $T_{\alpha\beta}(x)t^{\alpha}(x)t^{\beta}(x)$. Using this, we find that the energy-energy correlator, between the spacetime points $x$ and $y$, is given by\begin{equation}
\frac{N_{0000}(x,y)}{a^2(\eta_{x})a^2(\eta_{y})} \, .
\end{equation} 
From the expression of the noise kernel on constant time slices, we find that the energy-energy correlator, on late constant time sheets, has leading order behaviour of the form $\frac{2H^8\eta^6}{\pi^3(\Delta \vec{x})^6}\frac{(1 + \frac{m^2}{H^2})(\frac{m^3}{H^3})}{\mathrm{sinh}(\frac{2\pi m}{H})}$. Although there is an extra factor of $(H\eta)^4$ in front of the energy  correlator compared to the noise kernel and hence leading to an even faster decay at late times, the qualitative remarks, made above for the noise kernel, hold equally good for this invariant correlator as well. Let us now move on to discuss similar studies but now in general FRW settings and for massless spinor fields only.
\section{Behaviour of the noise kernel and the energy-energy correlator for massless spinors in general FRW spacetimes}
In this section, we analyse the behaviour of the noise kernel and the energy-energy correlator for massless spinor fields in general FRW spacetimes. To carry out this task, we make use of the equivalence of massless spinors in FRW spacetimes with massless spinors in de Sitter spacetime to relate the corresponding noise kernels. This equivalence between massless spinors in FRW and de Sitter spacetimes is analogous to the equivalence between massless scalar fields in FRW spacetimes with massive scalar fields in de Sitter spacetime established in \cite{Lochan:2018pzs}. Now recalling that a massless spinor field, $\psi$, in an FRW spacetime (with scale factor, $ c(\eta)$) is related to a massless spinor field, $\Omega$, in de Sitter spacetime by the relation $\psi(x)$ = $\Big(\frac{a(\eta)}{c(\eta)}\Big)^{\frac{3}{2}}\Omega(x)$ (see subsection $2.2$), we find that the Wightmann functions in the corresponding spacetimes are related by  
\begin{equation}
S^{FRW}_{ij}(x,y) =  \Big(\frac{a(\eta)}{c(\eta)}\Big)^{\frac{3}{2}}\Big(\frac{a(\eta')}{c(\eta')}\Big)^{\frac{3}{2}}S^{dS}_{ij}(x,y) 
\end{equation} 
and 
\begin{equation}
R^{FRW}_{ji}(y,x) =   \Big(\frac{a(\eta)}{c(\eta)}\Big)^{\frac{3}{2}}\Big(\frac{a(\eta')}{c(\eta')}\Big)^{\frac{3}{2}}R^{dS}_{ji}(y,x) \, ,
\end{equation}
where $a(\eta) = -\frac{1}{H\eta}.$ Using these expressions in equation (\ref{3}), we see that the considered component of the FRW noise kernel is given by\footnote{Here, like in the case of de Sitter scale factor ($a_{x}$), $c_{x}$ stands for $c(\eta_{x})$.}
\begin{multline}
N^{FRW}_{0000}(x,y) =  \lim_{x' \to x}\lim_{y' \to y}\frac{c_{x}c_{y}}{4}\Bigg[\partial^{x'}_{0}\partial^{y'}_{0} + \partial^{x}_{0}\partial^{y}_{0} - \partial^{x'}_{0}\partial^{y}_{0} - \partial^{x}_{0}\partial^{y'}_{0}\Bigg]H^4 \frac{(a_{x}a_{x'}a_{y}a_{y'})}{(c_{x}c_{x'}c_{y}c_{y'})^{\frac{3}{2}}}\\ Tr\Bigg(\Gamma_{0}\Big[i\Gamma^{\lambda}\partial^{x'}_{\lambda} + i\frac{ a'_{x'}}{a_{x'}}\Gamma^0\Big] \Big[\sum_{\epsilon=\pm}S_{\epsilon}(x',y)\frac{1 + \epsilon\Gamma^{0}}{2} \Big] \Gamma_{0}\Big[i\Gamma^{\sigma}\partial^{y'}_{\sigma} + i\frac{ a'_{y'}}{a_{y'}}\Gamma^0\Big] \Big[\sum_{\epsilon=\pm}S_{\epsilon}(y',x)\frac{1 + \epsilon\Gamma^{0}}{2} \Big]\Bigg) \, .
\end{multline}
After operating all the derivative operators (present in the bigger square brackets) on the $H^4 \frac{(a_{x}a_{x'}a_{y}a_{y'})}{(c_{x}c_{x'}c_{y}c_{y'})^{\frac{3}{2}}}$ term, we find that this term comes in front of the bigger square brackets as it is. Therefore, we can conclude that the massless spinor field noise kernel in general FRW spacetimes is just $\frac{(a_{x}a_{y})^2}{(c_{x}c_{y})^{2}}$ factor multiplying the de Sitter noise kernel expression of the previous section along with the fact that we also have $m = 0 $ and hence $S_{+}(Z(x,y)) = S_{-}(Z(x,y)) = \frac{1}{16\pi^2(1-Z(x,y))}.$ Like in the previous section, we consider the $(a=b=c=d=0)$ component of the noise kernel on equal time sheets i.e., $\eta_{x}= \eta_{y} = \eta$, which, in this case, is given by
\begin{equation} 
N^{FRW}_{0000}(x,y) = \frac{a_{x}^4}{c_{x}^{4}}\frac{3H^4\eta^4}{2\pi^4(\Delta \vec{x})^8}  = \frac{3}{2\pi^4(\Delta \vec{x})^8} c_{x}^{-4} \, .
\end{equation} 
Thus, we see that the considered component of the noise kernel of a massless spinor field in an FRW spacetime  (with scale factor being $c(\eta)$) behaves in a manner that is opposite to the behaviour of the scale factor. This implies that the correlations, between (massless) spinor matter located at spatially separated points on constant time sheets, decay during the expanding phases of the universe while they grow during the contracting phases of the universe. We also notice that the behaviour of the noise kernel component is monotonic (non-monotonic) if the scale factor changes monotonically (non-monotonically). By monotonic growth (or decay) of the noise kernel, we simply mean that the value of the noise kernel for spatially separated points on any constant time sheet is always more (or less) than its values on earlier time-sheets. We now specialize to the cosmologically interesting power-law type expanding FRW spacetimes i.e., $c(\eta)\propto \eta^{-q}$. For these cases, we have 
\begin{equation} 
N^{FRW}_{0000}(x,y)  = \frac{3}{2\pi^4(\Delta \vec{x})^8} (H\eta)^{4q} \, .
\end{equation} 
\\ \noindent Since $\eta \to 0$ limit is the late time limit for spacetimes with positive values of q, we notice that, for these spacetimes, the considered component of the noise kernel vanishes in this limit. For these spacetimes i.e., $q \in (0,\infty)$, the equation of state parameter, $w$, lies in the range $(-\infty,-\frac{1}{3})$ and these are acceleratingly expanding spacetimes. These spacetimes include both phantom cosmologies (see \cite{Johri:2003rh}) i.e., $w \in (-\infty,-1)$, as well as quintessence cosmologies i.e., $w \in (-1,-\frac{1}{3})$. In fact, the most accurate data till date puts the present day dark-energy driven universe in this regime i.e., $\omega = -1.03 \pm 0.03$ \cite{Planck:2018vyg}. For spacetimes with negative values of $q$, the late time limit is the $\eta \to \infty$ limit and we see that, for these spacetimes also, the considered component of the noise kernel vanishes in the late time limit. For these spacetimes i.e., $q \in (-\infty,0)$, the equation of state parameter, $w$, lies in the interval $(-\frac{1}{3},\infty)$. These spacetimes are also expanding spacetimes but, for them, the rate of expansion decreases with time i.e., the expansion is decelerating. These spacetimes include cosmologically interesting epochs like radiation dominated and dust dominated cases as well. These considerations imply that the vanishing of the noise kernel component, in the scaling factor going to large value limit, is universal for all power-law type expanding FRW cosmologies and we know that this should be the case for expanding spacetimes as we have already seen that the behaviour of the considered noise kernel component is opposite compared to the behaviour of the scaling factor. As argued in the de Sitter case, this implies that the quantum dynamics of massless spinor fields in power-law type FRW spacetimes is not such that it can overcome the effects of increasing physical distances between spatially separated points. Thus, we see that the quantum correlations between local stress-energy operators (at different points on constant time slices) are suppressed by the increasing physical distances in these expanding spacetimes. 
\\ Like in the previous section, we see that the considered noise kernel component may change its form under coordinate transformations. Therefore, we look for the coordinate independent quantity of the previous section i.e., the energy-energy correlator which is given by (on equal time sheets)
\begin{equation}
\frac{N^{FRW}_{0000}(x,y)}{c^2(\eta_{x})c^2(\eta_{y})} \, = c_{x}^{-8}\frac{3}{2\pi^4(\Delta \vec{x})^8}.
\end{equation}
This implies that, like the considered component of the noise kernel, the behaviour of the energy-energy correlator is also opposite to the behaviour of the scale factor (of the FRW spacetime). In particular, it vanishes for all power-law type expanding FRW spacetimes. All the remarks made above for the $(a=b=c=d=0)$ component of the noise kernel hold for the energy-energy correlator as well except for the fact that the energy-energy correlator depends on even more negative power of the scale factor compared to the considered noise kernel component. \\
For the special case of power-law type FRW spacetimes, one can compare these results for massless spinor fields with the corresponding results for massless scalar fields \cite{Dhanuka:2020yxp}. Whereas in the case of massless spinor fields in power-law FRW spacetimes, there are no 2nd order quantum corrections, \cite{Dhanuka:2020yxp} shows that, in case of massless scalar fields evolving in these FRW spacetimes, there are a number of these power-law FRW spacetimes for which the 2nd order quantum corrections coming from the noise kernel are significant e.g., for $\omega \in (0,-\frac{1}{3})\cup(-\frac{1}{3},-1)$ universes there are large quantum fluctuations. This implies that, for scenarios where we have both massless spinor and massless scalar fields present in power-law FRW spacetimes, the results, obtained by performing only the (first order) semiclassical analysis, get corrections only from the scalar sector but not from the spinor sector. 
 
\section{Summary and conclusions}
In this work, we have tried to understand the dynamics of arbitrarily massive spinor fields living on de Sitter spacetime. In particular, we have looked at the correlations between local stress energy operators in the late time limit while assuming the Bunch Davies initial conditions for the spinor fields at early times. We have also considered the behaviour of these correlations for masssless spinor fields in general FRW spacetimes. In order to carry out this analysis, we employed a conformal mapping from massless spinor fields in a general FRW spacetime to massless spinor fields in de Sitter spacetime. This mapping helps us place massless spinor fields (in FRW spacetimes) in Bunch Davies like vacua. Below, we give a brief summary of the results obtained in this work:  
\begin{itemize}
\item {\textbf{Behaviour of the noise kernel for spinor fields in de Sitter spacetime:} We consider the $(a=b=c=d=0)$ component of the noise kernel for a spinor field in the de Sitter spacetime. We find that, in the late time i.e., $\eta \to 0$ limit, the leading order behaviour tells us that the considered component of the noise kernel (on constant time slices) decays. From this, we infer that the quantum dynamics of the spinor fields is washed away by the classical accelerated expansion of the de Sitter spacetime. This dominance of the classical dynamics of the background over the quantum dynamics of the spinor field is universal, in the sense that it occurs irrespective of the mass of the field. 
This behaviour of the noise kernel for spinor fields is in stark contrast to the behaviour of the noise kernel for scalar fields (in de Sitter spacetime) \cite{Dhanuka:2020yxp} where the noise kernel, considered at late times, decays for scalar fields with ($2 <\frac{m^2}{H^2}< \frac{9}{4}$) but diverges for scalar fields with ($0 <\frac{m^2}{H^2}< 2$). 
For spinor fields, we observe that the invariant energy-energy correlator also decays. In fact, the decay is faster for this correlator. We also find that these correlators show the usual dependence on the coordinate distance for points on equal time sheets i.e., that these correlations become less and less important as the comoving distances increase. }
\item {\textbf{Behaviour of the noise kernel for spinor fields in FRW spacetimes:} Using the equivalence that exists between massless spinor fields in FRW spacetimes and massless spinor fields in de Sitter spacetime, we consider the behaviour of the $(a=b=c=d=0)$ component of the noise kernel for massless spinor fields in FRW spacetimes, again on constant time slices. In this setting, we find that the considered component of the noise kernel behaves in a manner that is opposite to the behaviour of the scale factor i.e., it decays (or grows) with expanding (or contracting) scale factors. This implies that the noise kernel component always decays for power-law type expanding FRW spacetimes. These results, for power-law type expanding FRW spacetimes, are in stark contrast to the behaviour of the considered noise kernel component for quantum scalar fields for which the noise kernel does not always decay \cite{Dhanuka:2020yxp}. Like the noise kernel component, the energy-energy correlator of the massless spinor fields is also inversely related to the scale factor.}
\end{itemize}
From the above results, one can conclude that, for systems involving massless spinor fields in power-law type expanding FRW spacetimes and arbitrarily massive spinor fields in de Sitter spacetime, the first order quantum treatments (based solely upon considering quantum averages of the stress energy operator) do not get significant corrections from the second order quantum effects coming from the noise kernel and hence, inferences made from a first order analysis will remain robust against the considered quantum fluctuations. 
It is important to emphasize that these conclusions have been arrived at for spinor fields by placing them in the Bunch Davies like vacua. 
Since the conclusions of field theories in curved spacetimes are markedly different for different vacua (\cite{Collins:2004wj,Goldstein:2002fc,Collins:2003ze}), it would be interesting to investigate how the noise kernel for spinor fields behaves for other de Sitter spacetime vacua like the fermionic alpha-vacua \cite{Collins:2004wj} etc. For general FRW spacetimes, we can again use the conformal invariance of massless spinor fields and carry out the analysis for the FRW vacua corresponding to other de Sitter space-time vacua.   
\section*{Acknowledgments}
AD would like to thank Kinjalk Lochan for carefully reading the manuscript of this paper and giving useful suggestions. AD would also like to acknowledge the financial support from University Grants Commission, Government of India, in the form of Senior Research Fellowship (UGC-CSIR JRF/Dec2016/510944). 
\section*{Appendix A}
Here we enlist some important formulae that are used in evaluating noise kernel in the main portion of the draft 
\begin{eqnarray}
Z(x',y)&=& 1 + \frac{(\eta_{x'}-\eta_{y})^2 - (\vec{x}'-\vec{y})^2}{4\eta_{x'}\eta_{y}}\\
\frac{\partial Z(x',y)}{\partial x'^{\mu}}&=& \frac{1}{2}\Bigg[\frac{\Delta s^2}{2\eta_{x'}^2\eta_{y}}\delta_{\mu 0} - \frac{(x'-y)_{\mu}}{\eta_{x'}\eta_{y}}\Bigg]\\
\frac{\partial Z(x',y)}{\partial y^{\nu}}&=& \frac{1}{2}\Bigg[\frac{\Delta s^2}{2\eta_{x'}\eta_{y}^2}\delta_{\nu 0} + \frac{(x'-y)_{\nu}}{\eta_{x'}\eta_{y}}\Bigg]\\
\frac{\partial^2 Z(x',y)}{\partial x'^{\nu}\partial x'^{\mu}}&=& \frac{1}{2}\Bigg[-\frac{\Delta s^2}{\eta_{x'}^3\eta_{y}}\delta_{\mu 0}\delta_{\nu 0} + \frac{(x'-y)_{\nu}}{\eta_{x'}^2\eta_{y}}\delta_{\mu 0}+ \frac{(x'-y)_{\mu}}{\eta_{x'}^2\eta_{y}}\delta_{\nu 0} - \frac{\eta_{\mu\nu}}{\eta_{x'}\eta_{y}}\Bigg]
\\
\frac{\partial^2 Z(x',y)}{\partial y^{\nu}\partial y^{\mu}}&=& \frac{1}{2}\Bigg[-\frac{\Delta s^2}{\eta_{x'}\eta_{y}^3}\delta_{\mu 0}\delta_{\nu 0} - \frac{(x'-y)_{\nu}}{\eta_{x'}\eta_{y}^2}\delta_{\mu 0} - \frac{(x'-y)_{\mu}}{\eta_{x'}\eta_{y}^2}\delta_{\nu 0} - \frac{\eta_{\mu\nu}}{\eta_{x'}\eta_{y}}\Bigg]
\\
\frac{\partial^2 Z(x',y)}{\partial y^{\nu}\partial x'^{\mu}}&=& \frac{1}{2}\Bigg[-\frac{\Delta s^2}{2\eta_{x'}^2\eta_{y}^2}\delta_{\mu 0}\delta_{\nu 0} - \frac{(x'-y)_{\nu}}{\eta_{x'}^2\eta_{y}}\delta_{\mu 0}+ \frac{(x'-y)_{\mu}}{\eta_{x'}\eta_{y}^2}\delta_{\nu 0} + \frac{\eta_{\mu\nu}}{\eta_{x'}\eta_{y}}\Bigg]
\\
\frac{\partial^3 Z(x',y)}{\partial y^{\rho}\partial x'^{\nu}\partial x'^{\mu}}&=& \frac{1}{2}\Bigg[\frac{\Delta s^2}{\eta_{x'}^3\eta_{y}^2}\delta_{\mu 0}\delta_{\nu 0}\delta_{\rho 0} - \frac{(x'-y)_{\nu}}{\eta_{x'}^2\eta_{y}^2}\delta_{\mu 0}\delta_{\rho 0} - \frac{(x'-y)_{\mu}}{\eta_{x'}^2\eta_{y}^2}\delta_{\nu 0}\delta_{\rho 0}  \\ & & + \frac{\eta_{\mu\nu}}{\eta_{x'}\eta_{y}^2}\delta_{\rho 0} -  \frac{\eta_{\mu\rho}}{\eta_{x'}^2\eta_{y}}\delta_{\nu 0} -  \frac{\eta_{\nu\rho}}{\eta_{x'}^2\eta_{y}}\delta_{\mu 0}   +2 \frac{(x'-y)_{\rho}}{\eta_{x'}^3\eta_{y}}\delta_{\mu 0}\delta_{\nu 0}
 \Bigg]
 \\
\frac{\partial^3 Z(x',y)}{\partial x'^{\rho}\partial y^{\nu}\partial y^{\mu}}&=& \frac{1}{2}\Bigg[\frac{\Delta s^2}{\eta_{x'}^2\eta_{y}^3}\delta_{\mu 0}\delta_{\nu 0}\delta_{\rho 0} + \frac{(x'-y)_{\nu}}{\eta_{x'}^2\eta_{y}^2}\delta_{\mu 0}\delta_{\rho 0} + \frac{(x'-y)_{\mu}}{\eta_{x'}^2\eta_{y}^2}\delta_{\nu 0}\delta_{\rho 0}  \\ & & + \frac{\eta_{\mu\nu}}{\eta_{x'}^2\eta_{y}}\delta_{\rho 0} -  \frac{\eta_{\mu\rho}}{\eta_{x'}\eta_{y}^2}\delta_{\nu 0} -  \frac{\eta_{\nu\rho}}{\eta_{x'}\eta_{y}^2}\delta_{\mu 0}   - 2 \frac{(x'-y)_{\rho}}{\eta_{x'}\eta_{y}^3}\delta_{\mu 0}\delta_{\nu 0}
 \Bigg]
\end{eqnarray}
%
\end{document}